\documentclass{main}

\usepackage{textcomp}
\usepackage{fancyhdr}
\usepackage{lastpage}
\usepackage{subcaption}
\usepackage[font=footnotesize]{caption}
\usepackage{wrapfig}
\usepackage{amsmath}
\usepackage{hyperref}
\usepackage{cleveref}
\usepackage{bm}
\usepackage{amssymb}
\usepackage[english]{babel}
\usepackage[autostyle, english = american]{csquotes}
\MakeOuterQuote{"}
\def\be{\begin{equation}}
\def\ee{\end{equation}}
\def\bea{\begin{eqnarray}}
\def\eea{\end{eqnarray}}

\fancypagestyle{firstpagefooter}{%
  \fancyfoot[L]{  \textcopyright \footnotesize This work is an open access article under a Creative Commons Attribution 4.0 International License (https://creativecommons.org/licenses/by/4.0/).}
  \fancyfoot[C]{}  
   
}

\begin{document}

\thispagestyle{firstpagefooter}
\title{\Large Describing UPC Data with the Sar$t$re Event Generator}

\author{Vaidehi Nattoja\footnote{email: vaidehinattoja@gmail.com} and Tobias Toll}

\address{
Indian Institute of Technology, New Delhi, [Delhi] 110016, India
}

\maketitle\abstracts{
Ultra-peripheral heavy-ion collisions (UPCs) provide a distinct environment for high-energy QCD research, focusing on the production of vector mesons. This proceeding details recent advancements in the Sar$t$re Monte Carlo event generator, a dipole model-based tool, to better describe UPC data. We present the incorporation of the full photon flux, accounting for interference effects and photon transverse momentum ($k_T$), and the integration of an $n_0^0n$ afterburner for neutron-tagged event classification. These extensions significantly improve Sar$t$re's ability to reproduce experimental data from STAR and LHCb, particularly at small $p_T^2$ and for various neutron event classes.
}



\section{Introduction}
UPCs are characterized by large impact parameters in hadronic interactions, which minimizes the strong interaction and allows electromagnetic forces to be the primary drivers. This allows for the study of photon-induced processes, such as exclusive vector meson production ($J/\psi$, $\rho$, $\phi$, $\Upsilon$), off heavy nuclei. The Sar$t$re event generator is a widely used Monte Carlo tool based on the dipole model, designed to simulate these processes. Previous versions of Sar$t$re successfully described exclusive $J/\psi$ production in STAR at large $p_T^2$ using the Hotspot Model. However, a comprehensive description of the full $p_T^2$ range and the various final states, particularly those involving forward neutron emission, required further theoretical developments. This work addresses these limitations by introducing the full photon flux, enabling the description of interference effects and the correct photon $k_T$, and by integrating an afterburner for neutron-tagged events.

\section{Theoretical Framework}
\subsection{Exclusive Diffraction in the Dipole Model}

\begin{figure}[htp]
    \centering
    \includegraphics[width=0.5\linewidth]{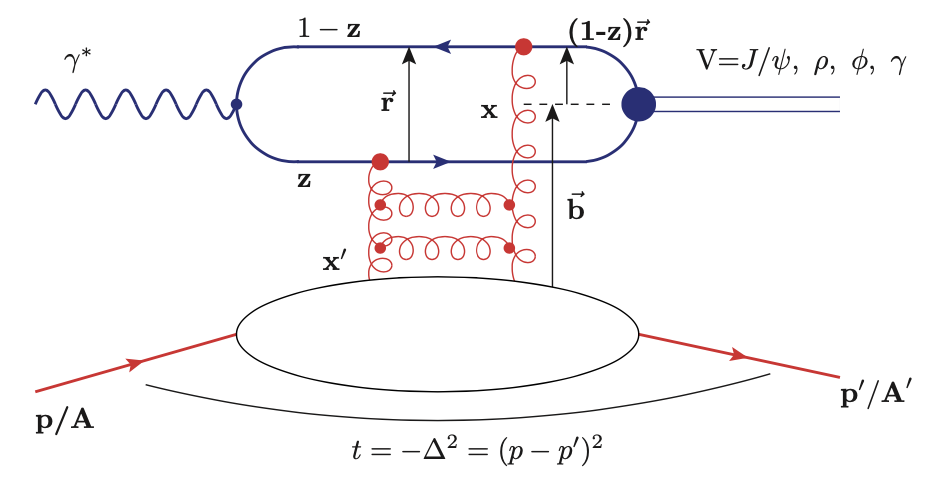} 
    \caption{Dipole Model}
    \label{fig:dipole}
\end{figure}

The dipole model conceptualizes deep inelastic scattering (DIS) as an interaction of a color dipole (primarily a $q\overline{q}$ pair) with the target proton\cite{Kowalski:2006hc} as shown in Fig.~\ref{fig:dipole}. It can be explained in the following three stages. The virtual photon spontaneously forms a color-neutral dipole $q\overline{q}$. This newly formed $q\overline{q}$ pair then scatters off the proton. The final stage is the recombination of the $q\overline{q}$ pair, which results in a vector meson ($J/\psi$, $\rho$, $\phi$, $\gamma$, etc).

The amplitude for exclusive vector meson production is given by:
\begin{equation}
\mathcal{A}_{T,L}^{\gamma^*p \to Ep}(x,Q,\Delta) = i\int \rm{d}^2\textbf{r}\int_{0}^{1}\frac{\rm{d}z}{4\pi}\int \rm{d}^2\textbf{b} \times (\Psi_V^*\Psi)_{T,L}e^{-i\left[\textbf{b}-\frac{(1-z)\textbf{r}}{2}\right]\cdot\Delta}N(\textbf{b},\textbf{r},x)
\end{equation}
where $(\Psi_V^*\Psi)$ represents the overlap between the vector meson and virtual photon wave functions, calculated from QED, and $N(b,r,x)$ is the dipole amplitude describing the scattering off the target\cite{Golec-Biernat:1999qor}. Two primary models for the dipole amplitude are considered\cite{Kowalski:2003hm}.

The \textbf{ bNonSat model } is a simpler linear model. It assumes that the interaction strength is directly proportional to the density of gluons in the proton. This is a good approximation for very small dipoles, where the probe is a point-like particle and the target gluons are sparse.

\begin{equation}
    N(b,r,x)=2F(x,r^2)T_p(b)
\end{equation}

The \textbf{bSat model} is a more complex, nonlinear model that accounts for gluon saturation. As the dipole size increases, it starts to "see" a higher density of gluons in the proton. Eventually, the gluons become so dense that they begin to overlap and screen each other. At this point, the interaction strength no longer increases linearly, but instead saturates or reaches a maximum. This model is essential for accurately describing the interaction of larger dipoles and the physics in a high-density environment.

\begin{align}
    N(b,r,x)&=2\left[1-\exp(-r^2F(x,r^2)T_p(b))\right] \\
    F(x,r^2)&=\frac{\pi^2}{2N_c}\alpha_s(\mu^2)xg(x,\mu^2)
\end{align}

Here, $T_p(b)=\frac{1}{2\pi B_p}\exp\left[-\frac{b^2}{2B_p}\right]$ is the Gaussian profile function of the proton. The scale $\mu^2=\mu_0^2+\frac{c}{r^2}$ determines where $\alpha_s$ and the gluon density are evaluated. The gluon density at the initial scale $\mu_0$ is parameterized as $xg(x,\mu_0^2)=A_g x^{-\lambda_g}(1-x)^{5.6}$. The parameters $A_g$, $\lambda_g$, C, and $\mu_0$ are determined through fits to DIS data.

\subsection{From \texorpdfstring{$ep$}{ep} to \texorpdfstring{$eA$}{eA} and the Hotspot Model}
To extend the framework from electron-proton ($ep$) to electron-nucleus ($eA$) collisions, the concept of independent scattering is employed. For the bSat model, the nuclear scattering amplitude is given by:
\begin{equation}
\frac{\rm{d}\sigma_{q\overline{q}}^{A}}{\rm{d}^{2}b}=2\left[1-\exp\left(-\frac{\pi^{2}}{2N_{c}}r^{2}\alpha_s(\mu^2)xg(x,\mu^2)\sum_{i=1}^{A}T(|b-b_{i}|)\right)\right]
\end{equation}
where $b_i$ assumes the Woods-Saxon distribution.

The \textbf{Hotspot Model}\cite{Mantysaari:2016jaz} provides a substructure of the nucleon geometry, particularly for protons:
\begin{equation}
T_p(b)\rightarrow\frac{1}{N_q}\sum_{i=1}^{N_q}T_q(b-b_i)
\end{equation}
where $T_q(b)=\frac{1}{2\pi B_q}\exp\left[-\frac{b^2}{2B_q}\right]$. Here, $b_i$ represents the transverse location of hotspots sampled from a Gaussian distribution with width $B_{qc}$, and $B_q$ is the width of these individual hotspots. $N_q$ is the number of hotspots. This model allows for variations in the proton configuration based on the choices of $B_{qc}$ and $B_q$.

\subsection{Good-Walker Picture and Incoherent Scattering}
The Good-Walker picture accounts for nuclear breakup. When the nucleus breaks up, the scattering is termed incoherent; when it remains intact, it's coherent. The total cross-section is the sum of coherent and incoherent contributions:

\begin{align}
\frac{\rm{d}\sigma_{\rm{coherent}}}{\rm{d}t}&=\frac{1}{16\pi}|\langle\mathcal{A}\rangle_{\Omega}|^{2} \\
\frac{\rm{d}\sigma_{\rm{total}}}{\rm{d}t}&=\frac{1}{16\pi}\langle|\mathcal{A}|^{2}\rangle_{\Omega}
\end{align}

The incoherent cross-section, representing events where the nucleus changes its internal state, is given by:
\begin{equation}
\sigma_{\rm{incoherent}} = \langle|\mathcal{A}|^{2}\rangle_{\Omega}-|\langle\mathcal{A}\rangle_{\Omega}|^{2}
\end{equation}
This expression effectively serves as the variance for a Gaussian geometry and is crucial for describing the diffractive pattern at larger $|t|$ \cite{Good:1960ba}.

\subsection{From \texorpdfstring{$eA$}{eA} to UPCs: Photon Flux and Interference}
In UPCs, the cross section for a photonuclear interaction is determined by the convolution of the photon flux and the photonuclear cross section:
\begin{equation}
\sigma_{X}=\int(\text{Photon flux})\times\sigma_{\gamma X}(E_{\gamma})\rm{d}E_{\gamma}
\label{eq:csInUPC}
\end{equation}
The photon flux as a function of photon energy is:
\begin{equation}
\frac{\rm{d}N_{\gamma}}{\rm{d}E_{\gamma}}=\int \rm{d}^{2}\textbf{b}P_{NOHAD}(\textbf{b})N(E_{\gamma},\textbf{b})
\end{equation}
where $N(E_{\gamma},b)=\frac{Z^{2}\alpha_{EM}}{\pi^{2}}\frac{E_{\gamma}}{\gamma^{2}}\left(K_{1}^{2}(E_{\gamma}b/\gamma)+\frac{1}{\gamma^{2}}K_{0}^{2}(E_{\gamma}b/\gamma)\right)$.


A critical aspect of this work is the inclusion of the full photon flux, which introduces interference effects and correctly accounts for the photon transverse momentum ($k_T$)\cite{Mantysaari:2022sux}. The differential cross section for exclusive vector meson production in UPCs is given by:
\begin{equation}
\frac{\rm{d}\sigma^{A_1+A_2 \to V_1+A_1+A_2}}{\rm{d}\textbf{p}^2\rm{d}y}=\frac{1}{4\pi}\int_{|\textbf{B}|>B_{min}}\rm{d}^{2}\textbf{B}|\langle\mathcal{M}^{j}(y,\textbf{p},\textbf{B})\rangle_{\Omega}|^{2}
\end{equation}
where
\begin{equation}
\langle\mathcal{M}^{j}(y,\textbf{p},\textbf{B})\rangle_{\Omega}=\int \frac{\rm{d}^{2}\textbf{k}}{(2\pi)^{2}}\langle\mathcal{A}(y,\textbf{p}-\textbf{k})\rangle_{\Omega}\mathcal{F}^{j}(y,\textbf{k})e^{-i\textbf{B}\cdot \textbf{k}}+\int \frac{\rm{d}^{2} \bm{\Delta}}{(2\pi)^{2}}\langle\mathcal{A}(-y,\bm{\Delta})\rangle_{\Omega}\mathcal{F}^{j}(-y,\textbf{p}-\bm{\Delta})e^{-i\textbf{B}\cdot\bm{\Delta}}
\end{equation}

\begin{figure}[h]
\begin{subfigure}{0.5\textwidth}
\centering
\includegraphics[width=0.8\linewidth]{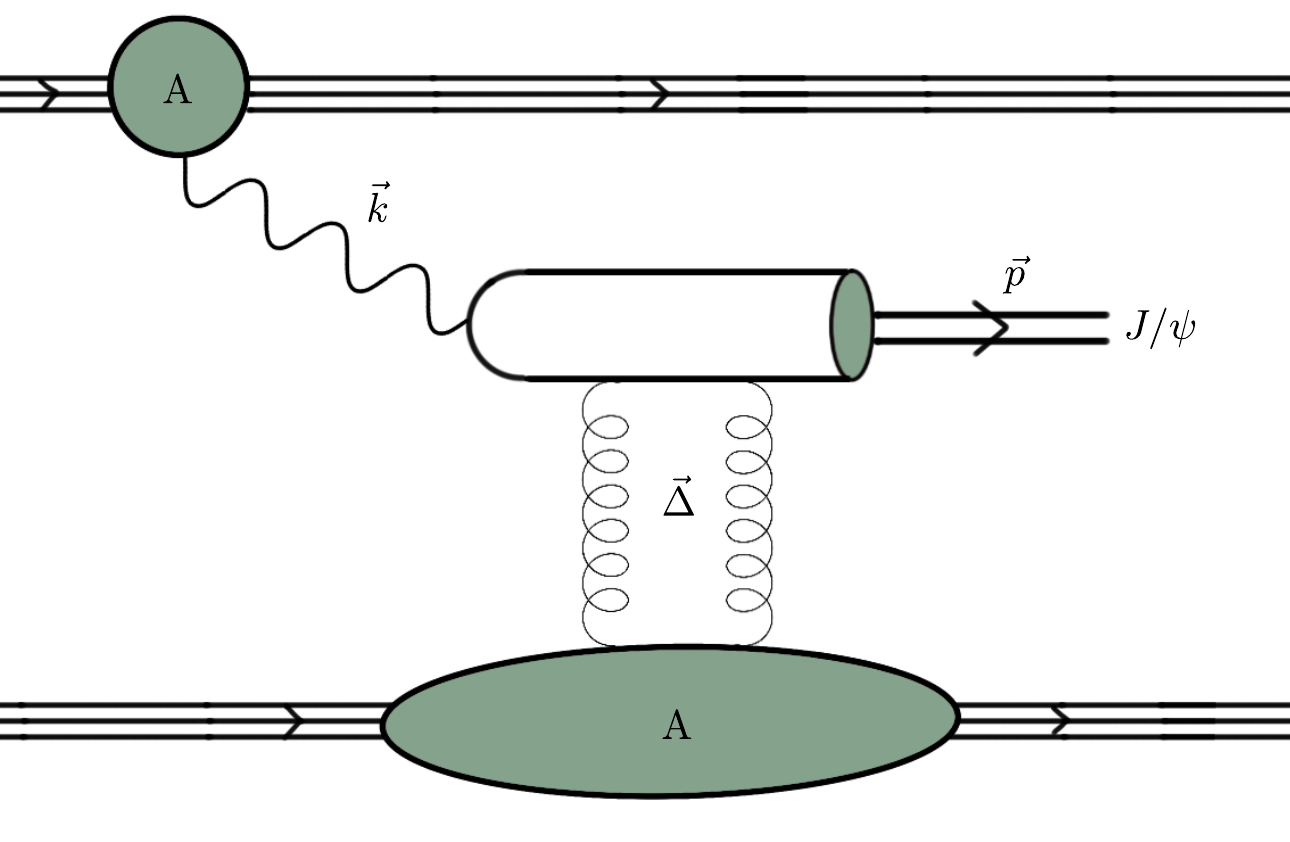} 
\label{fig:fd1}
\end{subfigure}
\begin{subfigure}{0.5\textwidth}
\centering
\includegraphics[width=0.8\linewidth]{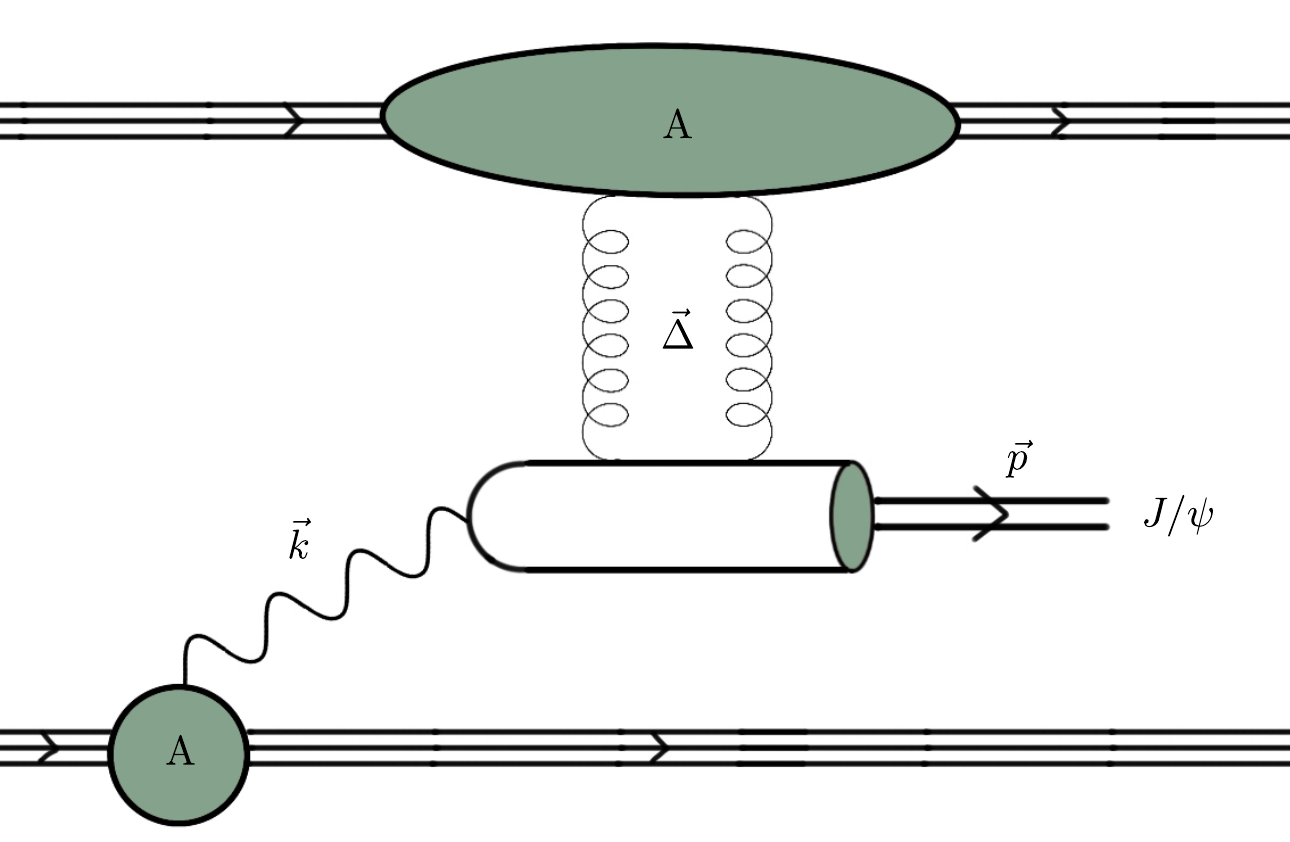}
\label{fig:fd2}
\end{subfigure}
\caption{Feynman diagrams illustrating vector meson photoproduction in UPCs, including photon flux and interference contributions.}
\label{fig:fd}
\end{figure}

Fig.~\ref{fig:fd} shows the underlying mechanism of vector meson photoproduction in UPCs. The two diagrams represent photon emission from either nucleus, whose amplitudes interfere. The convolution of the photon flux with the production amplitude, evaluated using Fast Fourier Transforms (FFTs), ensures that both the photon transverse momentum ($k_T$) and the interference pattern are correctly incorporated into the cross-section description.

Here, $\mathcal{F}^{j}$ is the photon field in the transverse plane, $\mathcal{A}$ is the vector meson production amplitude, $k$ is the transverse momentum of the photon, and $\Delta$ is the momentum transfer from the gluon field. Fast Fourier Transforms (FFTs) are applied to both terms in the amplitude. The incoherent cross section remains untouched in this formulation. The photon field $\mathcal{F}^{j}(y,k)$ is given by:
\begin{equation}
\mathcal{F}^{j}(y,\textbf{k})=2Z{\alpha_{em}}^{1/2}\textbf{k}^{j}\left[\frac{F(\textbf{k}^{2}+x_{\mathbb{P}}^{2}M_{p}^{2})}{\textbf{k}^{2}+x_{\mathbb{P}}^{2}M_{p}^{2}}\right]
\end{equation}
where $F(k^{2})=\frac{4\pi\rho^{0}}{k^{3}A}\left(\frac{1}{a^{2}k^{2}+1}\right)\left[\sin(kR_{A})-kR_{A}\cos(kR_{A})\right]$\cite{Klein:1999qj}.

\subsection{Integrating Sartre with \texorpdfstring{$n_0^0n$}{n00n} for Neutron Tagging}
To describe neutron-tagged event classes, Sar$t$re has been supplemented with the $n_0^0n$ Monte Carlo generator\cite{Broz:2019kpl}, which simulates forward neutron production by Coulomb excitation. This integration allows for a more detailed classification of events based on the number of emitted neutrons. The cross section for such processes is:


\begin{equation}
\sigma(AA\rightarrow PA_{i}^{\prime}A_{j}^{\prime})\propto\int \rm{d}^{2}b~P_{P}(b)P_{ij}(b)e^{-P_{H}(b)}
\end{equation}
where $P_H(b)$ is the probability of hadronic interaction, and $P_P(b)$ is the probability of the hard photoproduction process, suppressed in UPCs. $P_P(b)$ is defined as:
\begin{equation}
P_P(b)=\int \rm{d}k\frac{\rm{d}^{3}n(\textbf{{b}},k)}{\rm{d}k~\rm{d}^{2}\textbf{b}}\sigma_{\gamma A\rightarrow PA}(k)
\end{equation}
The probability of nuclear breakup is assumed to be independent, $P_{ij}(b)=P_i(b)\times P_j(b)$. The probability of observing exactly $L$ neutrons follows a Poisson distribution:
\begin{equation}
P_{n}^{L}(b)=\frac{(P_{Xn}^{1}(b))^{L}}{L!}e^{-P_{X_{n}}^{1}(b)}
\end{equation}
where $P_{X_n}^1(b)$ is the mean number of Coulomb excitations and $\sigma_{\gamma A\rightarrow A^{\prime}+X_n}(k)$ is determined from experiments on fixed targets with energies of up to 16.4 GeV.

\begin{figure}[h]
\centering
\includegraphics[width=0.5\linewidth]{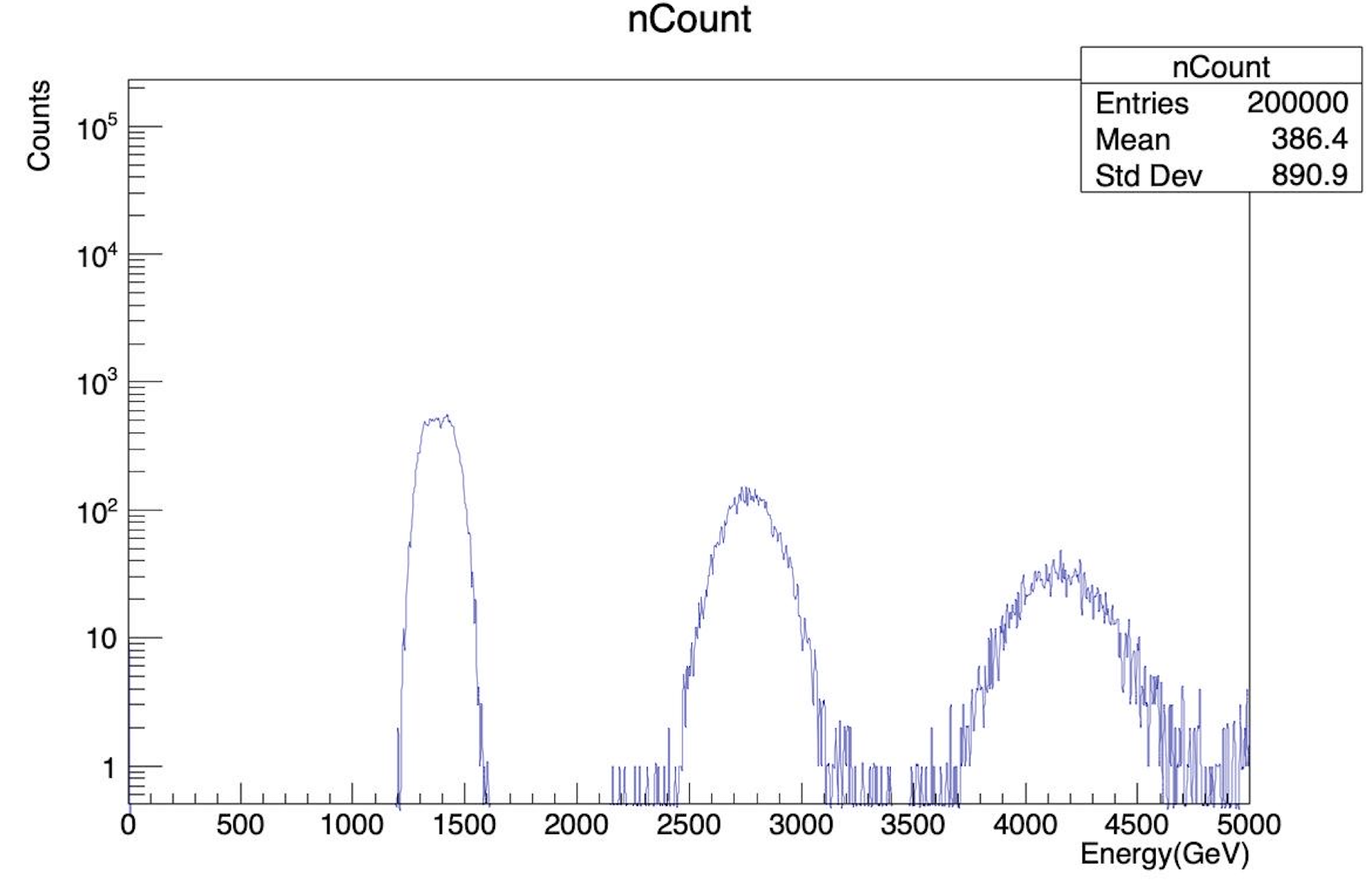} 
\label{fig:noon}
\caption{Emitted neutron count generated using $n_0^0n$ in Sar$t$re}
\end{figure}

\section{Experimental Comparisons and Results}
The extended Sar$t$re generator has been compared with the experimental data of STAR and LHCb for exclusive $J/\psi$ production.

\subsection{STAR Data Comparison}
Previous versions of Sar$t$re, incorporating sub-nuclear fluctuations (Hotspot Model), described the large $p_T^2$ region of STAR data well. However, they lacked the interference term and proper photon $k_T$ treatment, leading to discrepancies at small $p_T^2$.
With the inclusion of the full photon flux, the updated Sar$t$re now successfully fills the first dip observed in the STAR data at a small $p_T^2$ and provides a much better description across the entire $p_T^2$ range. This is evident in the comparison of the differential cross section $d^2\sigma/dp_T^2 dy$ versus $p_T^2$. The coherent and incoherent contributions, along with the interference term, are now accurately modeled for large and small $p_T^2$. However it overshoots for inermediate $p_T^2$.

\subsection{LHCb Data Comparison}
The updated Sar$t$re also shows good agreement with LHCb data for exclusive $J/\psi$ production. The model accurately reproduces the shape and magnitude of the cross section, further validating the importance of including interference effects and the full photon $k_T$. The characteristic dip observed in the coherent cross section is well described.


\begin{figure}[ht]
\begin{subfigure}{0.6\textwidth}
\centering
\includegraphics[width=0.9\linewidth]{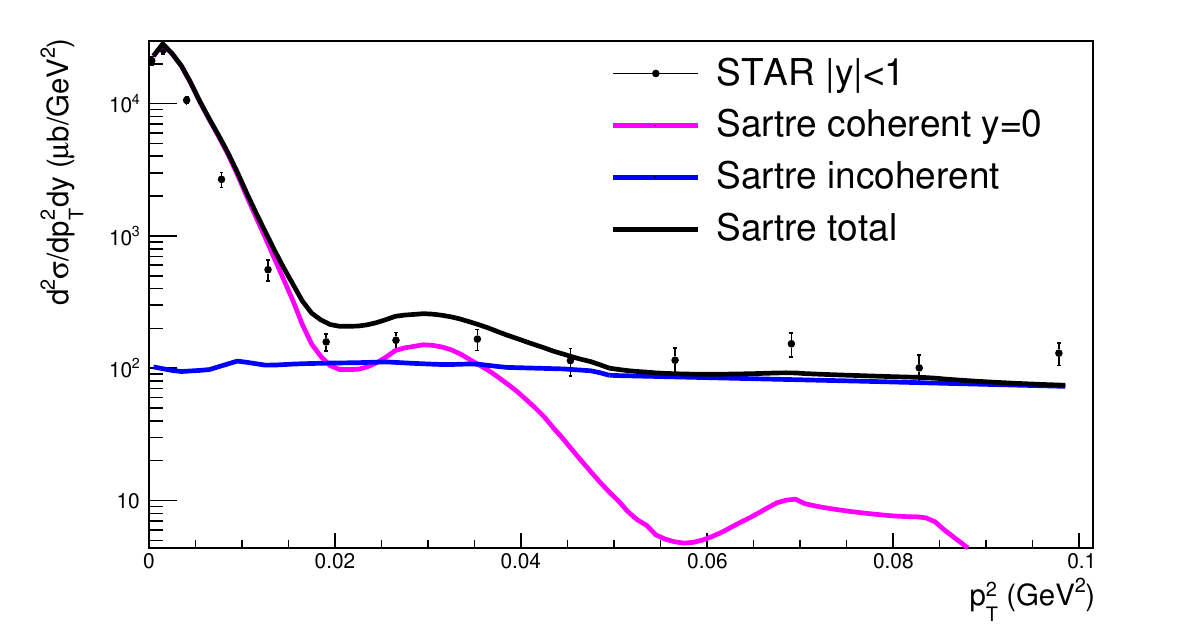} 
\caption{$\rm{Au}+\rm{Au} \rightarrow  \rm{Au}+\rm{Au} +\rm{J}/\psi$ at STAR}
\end{subfigure}
\begin{subfigure}{0.4\textwidth}
\centering
\includegraphics[width=0.9\linewidth]{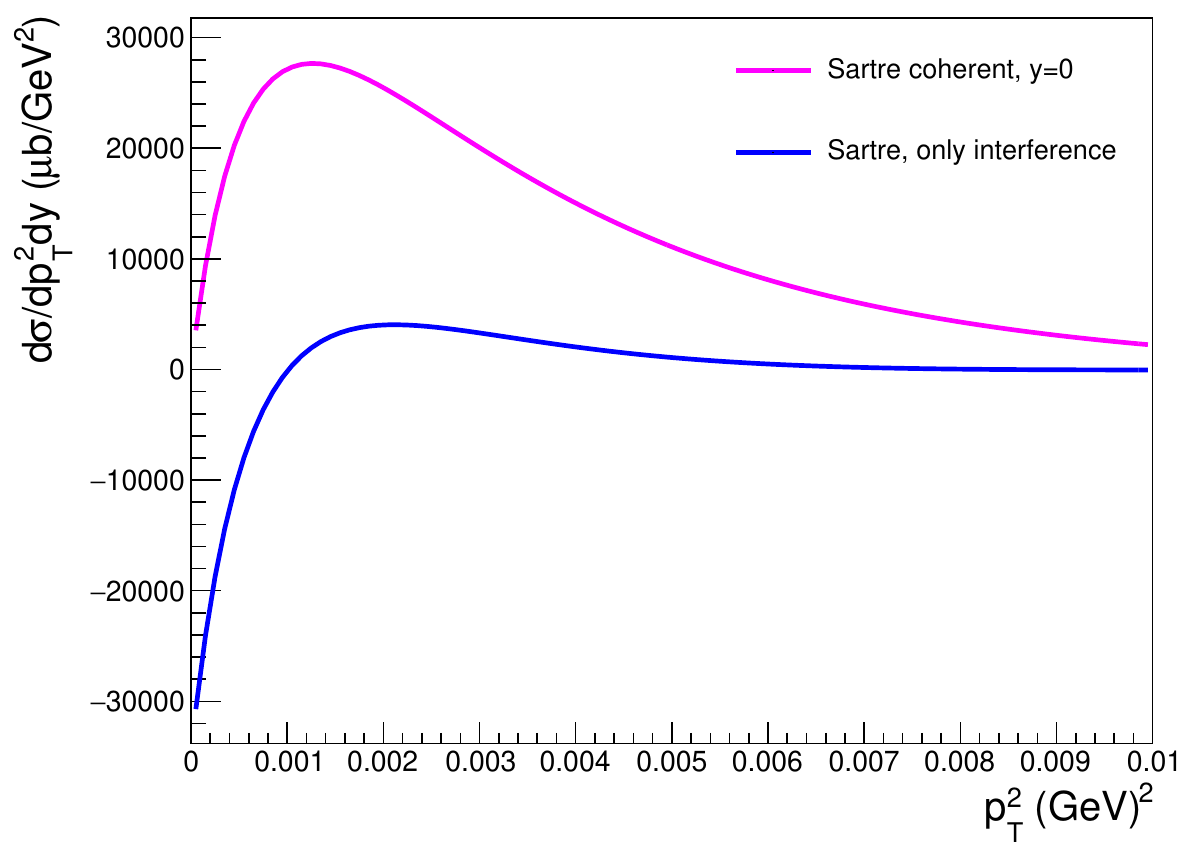}
\caption{Interference term only}
\end{subfigure}
\caption{Comparison of Sartre predictions with STAR data, showing coherent, incoherent, and interference contributions to exclusive $J/\psi$ photoproduction}
\label{fig:result}
\end{figure}

\section{Summary and Outlook}
The Sar$t$re event generator has been significantly improved to provide a more comprehensive description of exclusive verctor meson production in ultraperipheral heavy-ion collisions. Key advancements include the integration of the complete photon flux, which now correctly includes interference and photon transverse momentum effects ($k_T$) as seen in Fig.~\ref{fig:result}. This was essential for accurately modeling the initial dip in STAR's data and describing the low $p_T^2$ region for both STAR and LHCb experiments. We have also integrated $n_0^0n$ as an afterburner, enabling the description of different neutron event classes, is vital for understanding the full experimental landscape of UPCs.

These improvements enhance Sar$t$re's predictive power and its ability to serve as a robust tool for analyzing UPC data.

\textbf{Near Future Plans:}

One of the main goals for the near future is to create a more efficient event generator that can simulate the production of various heavy vector mesons, specifically $J/\psi$, $\phi$, and $\Upsilon$. Additionally, the plan is to expand the scope of Sartre to include the physics of inclusive diffraction for heavy flavors in UPCs, which will further broaden its application and enhance its ability to describe a wider range of high-energy collision phenomena.

These ongoing developments aim to make Sar$t$re an even more suitable tool for probing the fundamental properties of QCD in the unique environment of ultraperipheral collisions.

\section*{References}
\bibliography{ref}

\end{document}